\documentclass[journal=jpccck,manuscript=article]{achemso}
\usepackage{amsmath,color}
\usepackage{graphicx}
\usepackage{amssymb}
\usepackage{caption}
\usepackage{subcaption}
\usepackage{rotating}

\author{Sandeep K. Jain}
\affiliation[Institute for Theoretical Physics,Universiteit Utrecht]
{Institute for Theoretical Physics, Universiteit Utrecht,
Princetonplein 5, 3584 CC Utrecht, The Netherlands}
\email{S.K.Jain@uu.nl}

\author{Vladimir Juri\v ci\' c}
\affiliation[Nordita and Stockholm University ]
{Nordita, Center for Quantum Materials, KTH Royal Institute of Technology and
Stockholm University, Roslagstullsbacken 23, S-106 91 Stockholm, Sweden}
\email{juricic@nordita.org}

\author{Gerard T. Barkema}
\affiliation[Department of Information and Computing Science, Universiteit Utrecht]
{Department of Information and Computing Science, Universiteit Utrecht,
Princetonplein 5, 3584 CC Utrecht, The Netherlands}

\title{Probing the Shape of a Graphene Nanobubble}
\keywords{Graphene, nanobubble, substrate, van der Waals pressure, Vibrational density of states}

\begin{document}


\begin{abstract}
Gas molecules trapped between graphene and various substrates in the form of bubbles are observed experimentally. The study of these bubbles is useful in determining the elastic and mechanical properties of graphene, adhesion energy between graphene and substrate, and manipulating the electronic properties via strain engineering. In our numerical simulations, we use a simple description of elastic potential and adhesion energy to show that for small gas bubbles ($\sim 10$ nm) the van der Waals pressure is in the order of 1 GPa. These bubbles show universal shape behavior irrespective of their size, as observed in recent experiments. With our results the shape and volume of the trapped gas can be determined via the vibrational density of states (VDOS) using experimental techniques such as inelastic tunneling and inelastic neutron scattering. The elastic energy distribution in the graphene layer which traps the nanobubble is homogeneous apart from its edge, but the strain depends on the bubble size thus variation in bubble size allows control of the electronic and optical properties.
\end{abstract}

Graphene has shown remarkable and unusual electronic and mechanical properties.\cite{CastroNeto2009,Geim2009} In particular, the latter have recently attracted a great deal of attention. \cite{Nair2012,Smith2013,Dolleman2016,Cartamil-Bueno2016} This is so because graphene has emerged as a very strong material.\cite{Lee2008} N-doped graphene has been recently reported as the stiffest material ever found.\cite{Milowska2013} Its mechanical properties are also fundamentally and practically interesting when it is in contact with other materials. For instance, due to a strong van der Waals force, graphene can build bilayer, twisted bilayer, trilayer and other heterostructures.\cite{Geim2013} These van der Waals heterostructures are useful in nanotechnology and semiconductor industry because of the flexibility to tailor their properties.

Yet another important physical and chemical aspect of graphene is its ability to trap gas molecules under high pressure when placed on different substrates, leading to a formation of bubbles with nanometer to micrometer sizes. Such nanobubbles have been experimentally observed in a graphene membrane placed on top of a SiO$_2$/Si substrate, \cite{Stolyarova2009,Georgiou2011} epitaxial graphene grown on 4H-SiC, \cite{BenGouiderTrabelsi2014} and in an irradiated graphene sample on Ir.\cite{Zamborlini2015} Experimentally, it is observed that absorbed water and hydrocarbons between graphene and substrates can also lead to sub-micron sized bubbles. \cite{Haigh2012,Kretinin2014} The bubbles have been used for Raman characterization of strained graphene \cite{Zabel2012} and have been reported to induce pseudo-magnetic fields greater than $300$ T at room temperature.\cite{Levy2010} The van der Waals pressure inside the bubble has an important role in determining the properties of the trapped materials.\cite{Vasu2016,Jiao2016} Pressure due to the confinement can modify the properties of a material, e.g., ice in graphene nanocapillaries at room temperature,\cite{Xu2010,Algara-Siller2015} nanocrystals and biological molecules trapped in graphene liquid cells. \cite{Yuk2012,Park2015,Wojcik2015}

Nanobubbles proved to be an effective tool for strain engineering which is used to modify electronic and mechanical properties of graphene.\cite{Lu2012,Neek-Amal2012} Controllable curvature of graphene bubbles can be used as optical lenses with variable focal length.\cite{Georgiou2011} Gas molecules in graphene bubbles on patterned substrates have shown remarkable impermeability \cite{Bunch2008} therefore suggesting the applicability in gas storage devices.\cite{Stolyarova2009,Leenaerts2008} Graphene nanobubbles are also used to measure elastic properties and adhesion energy between substrates.\cite{Koenig2011} Very recently, nanobubbles of different sizes, ranging from a few tens of nanometers to a micrometer, and shapes, such as circular, trapezoidal, and triangular, have been found in van der Waals heterostructures.\cite{Khestanova2016} Various shaped small nanobubbles formed due to the trapping of an inert gas (Ar) between the graphene and the Ir substrate have been studied experimentally; control of the bubble's size by means of thermal treatment is suggested.\cite{Zamborlini2015} Importantly, circular bubbles show universal scaling behavior which can be captured by elastic continuum theory.\cite{Khestanova2016} However, it is not known whether this universal behavior extrapolates to the bubbles of smaller sizes down to $\sim 10$ nm.  

In this letter, we numerically simulate graphene nanobubbles of various sizes, starting with the ones with a radius ($R$) of $\sim 10$ nm. We use a computationally cheap and accurate semi-empirical potential to simulate the graphene sheet. The adhesion is modeled by a simple potential that produces essential elastic properties of the bubble, such as universal scaling of the aspect ratio $h/R$, with $h$ as its height, observed in large circular bubbles. Furthermore, we find that the pressure of trapped ideal gas for the smallest bubbles of size $\sim 10$ nm shows scaling with the aspect ratio as the elastic continuum theory predicts. The van der Waals pressure for these bubbles is in the order of 1 GPa. We find that the vibrational density of states (VDOS) can be used as an independent tool to detect the volume of gas trapped inside the bubble. It is interesting to notice that many soft vibrational modes appear in the VDOS because of the formation of bubbles. Finally, we map out the profile of elastic energy, as this can be used to extract the strain distribution of the bubble.

To simulate the monolayer graphene, we use a recently developed effective semiempirical elastic potential \cite{Sandeep-2015}, which was shown to effectively capture the structural properties of twisted bilayer graphene \cite{Jain2016} given by
\begin{align}\label{eff-potential}
E&=\frac{3}{16}\frac{\alpha}{d^2}\sum_{i,j}(r_{ij}^2-d^2)^2
  +\frac{3}{8}\beta d^2\sum_{j,i,k}\left(\theta_{jik}-\frac{2\pi}{3}\right)^2
  +\gamma\sum_{i,jkl}r_{i,jkl}^2.
\end{align}
Here,
$r_{ij}$ is the distance between two bonded carbon atoms, $\theta_{jik}$ is the angle between the two bonds connecting atom $i$ to atoms $j$ and $k$, and $r_{i,jkl}$ is the distance between atom $i$ and the plane through the three atoms $j$, $k$ and $l$ connected to atom $i$. The parameter $\alpha=26.060$~eV/\AA$^{2}$ controls bond-stretching and is fitted to the bulk modulus, $\beta=5.511$~eV/\AA$^{2}$ controls bond-shearing and is fitted to the shear modulus, $\gamma=0.517$~eV/\AA$^{2}$ describes the stability of the graphene sheet against buckling, and $d = 1.420$ \AA~is the ideal bond length for graphene. Additionally, the interaction between substrate and graphene layer is described by an extra harmonic term
\begin{equation}\label{eq-substrate}
E_S=K\sum_{i}\frac{z_i^2}{1+({z_i}/{z_0})^2},
\end{equation}
where $K$ is the effective elastic constant for the graphene-substrate interaction which determines the strength of the adhesion of graphene to the substrate, $z_0$ is a constant which sets the range of the harmonic regime of the energy term, and $z_i$ is the distance of carbon atom $i$ from the graphene plane. The theoretical prediction of adhesion energy for various substrates is $\sim 0.01-0.02$~eV/\AA$^{2}$, ~\cite{Bjorkman2012,Sachs2011} and  therefore in our numerical simulations we use the values of elastic constant $K=0.01$~eV/\AA$^{2}$ and $z_0=2$~\AA~ to capture the effects of various substrates.

\begin{figure}
\centering
\includegraphics[width=1.0\textwidth]{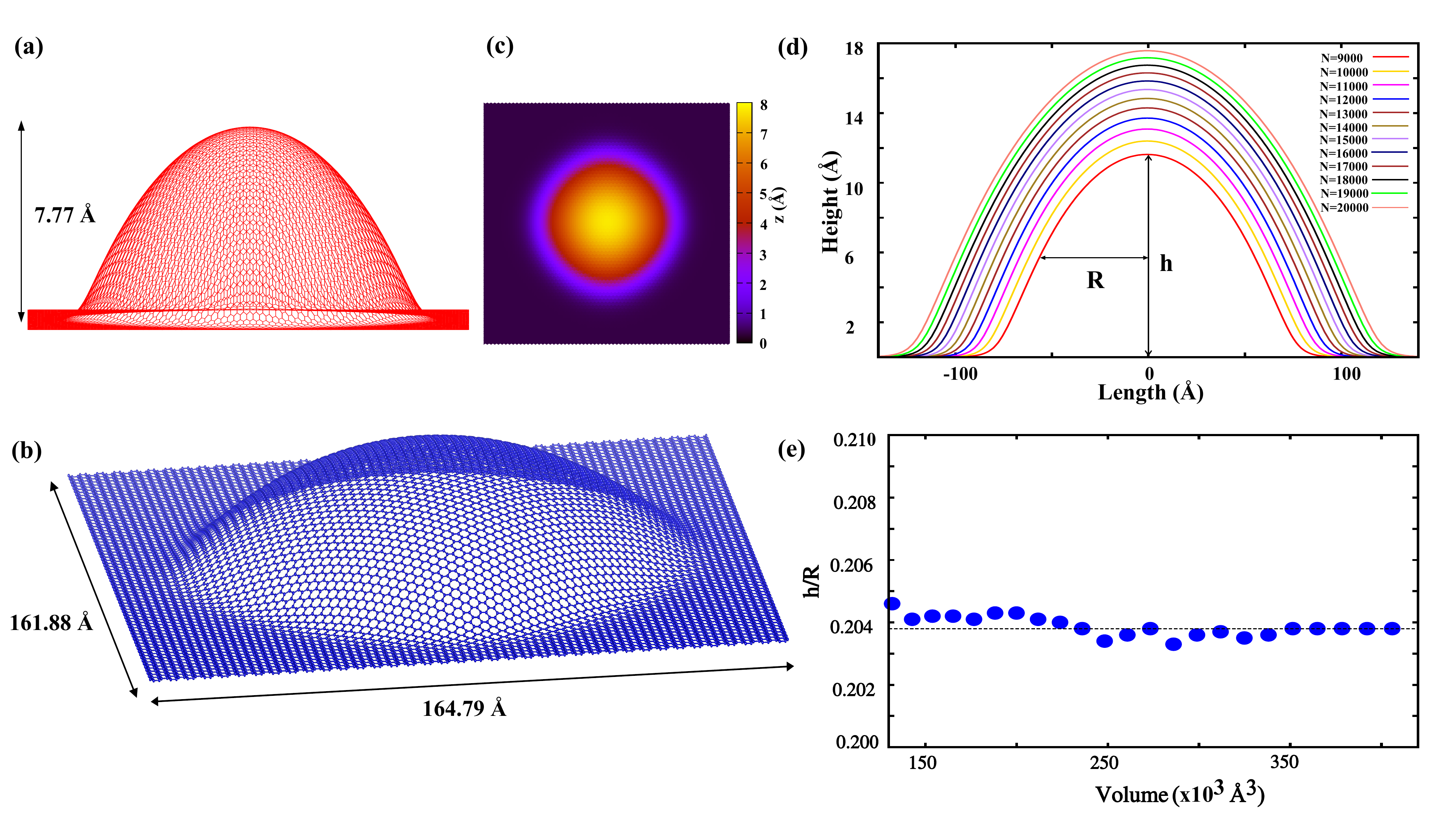}
\caption{Shape and profile of a gas bubble at the interface between graphene and a substrate. (a) Side view of a bubble trapped in a graphene layer with 10184 carbon atoms. The height ($h$) of the bubble is 7.77\AA. For better visibility the height scale is increased. (b)  Top view of the same bubble with its lateral dimensions. (c) Height profile of the same bubble. (d) 2D profile of the bubbles with varying number of the ideal gas molecules ($N$) trapped under a graphene layer with 34160 carbon atoms. The radius ($R$) of the bubble is measured at half of the bubble's maximum height. (d) Measured aspect ratios ($h/R$) as a function of volume show the constant behavior implying the universal shape of the bubble.}
\label{Fig-1}
\end{figure}

In our numerical simulations, a cone shaped initial void is created by pulling some carbon atoms out of the graphene plane and a fixed number of ideal gas molecules ($N$) are placed in the void. The whole system is then relaxed by above described potential (eq.\ (\ref{eff-potential})) combined with the substrate potential ($E_S$) (eq.\ (\ref{eq-substrate})) and the ideal gas law. We use deformation free (DF) periodic boundary conditions in our simulations.\cite{Jain2016a} The resulting structure has the well relaxed round shaped bubble filled with gas molecules as shown in Fig.\ \ref{Fig-1}. The vibrational spectrum is obtained by diagonalising the hessian matrix generated from above potential.\cite{Jain2015} In our plot, the VDOS is convoluted with a Gaussian function with a width of $\sigma=14$ cm$^{-1}$.


The graphene bubble we obtain using numerical simulations  is shown in Figs.\ \ref{Fig-1}(a) (side view) and \ref{Fig-1}(b) (top view). The round-like shape of the bubble results from the competition of the forces arising from pressure exerted by the ideal gas on the membrane and the elastic forces of graphene and substrate encoded in the semi-empirical potential (1) in our simulations. The height of the bubble, shown in Fig. \ref{Fig-1}(c), varies with the radius of the bubble, see Fig.\ \ref{Fig-1}(d). The aspect ratio $h/R\simeq0.204$ remains constant as the volume of the bubble changes with the number of the trapped molecules of the ideal gas, as shown in Fig.\ \ref{Fig-1}(e). This ratio only depends on the elastic properties of graphene and van der Waals (vdW) attraction between graphene and substrate, and is therefore independent of the properties of the trapped substance. This result is in excellent agreement with the prediction from the elastic continuum theory \cite{Khestanova2016} and continuum membrane plate theory.\cite{Wang2013,Yue2012} Furthermore, it demonstrates that the elastic continuum can be used for the description of bubbles of much smaller sizes than the ones experimentally studied by Khestanova \textit{et. al}.\cite{Khestanova2016} On the other hand, this implies that the deviations from this universal ratio are caused by external effects (i.e., residual strains in the samples, experimental conditions), which provides a possible route for the control of the shape of the bubble. We also study the universal shape behavior in much bigger gas bubbles trapped in a graphene sheet in simulations with 441504 carbon atoms as shown in the supplementary online material (Fig. S1).

\begin{figure*}
\centering
\includegraphics[width=0.5\textwidth]{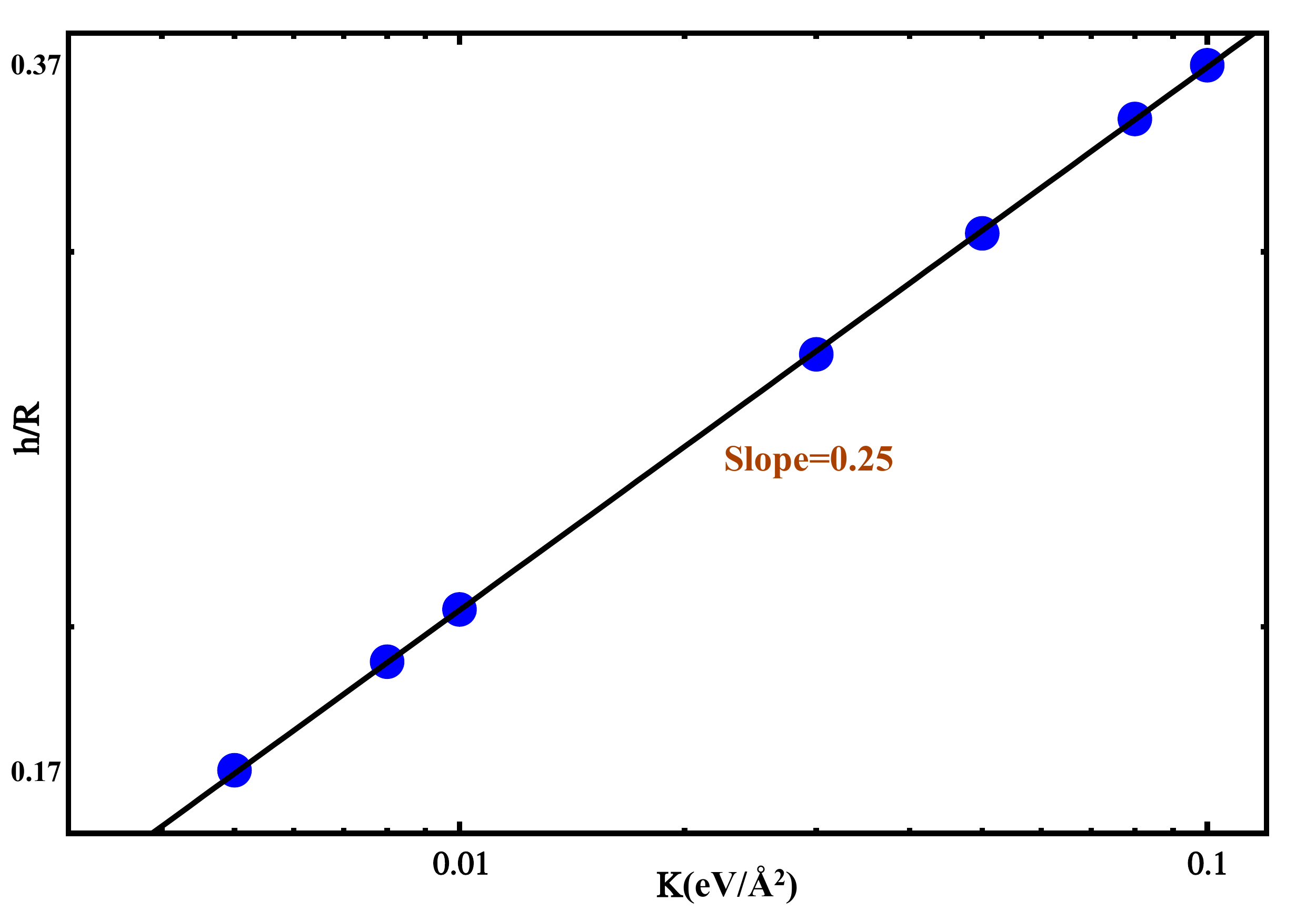}
\caption{Log-log plot between aspect ratio $h/R$ and the intensity $K$ of the substrate potential in eq.\ (\ref{eq-substrate}). The numerical data points (blue) are fitted with a straight line with a slope of $0.25$. This shows that the aspect ratio scales as $h/R \sim K^{1/4}$.}
\label{Fig-2}
\end{figure*}

To further theoretically demonstrate the predicted universal properties of the graphene bubble, we study the dependence of the aspect ratio on the strength of the substrate potential, $K$, which in our simulations plays the role of the adhesion energy in the experiments. Here, we vary the strength of this potential, which is not possible in experiments without changing substrates, and allow the bubble to relax.  We observe the scaling of the aspect ratio with the strength of the substrate potential, $h/R\sim K^{1/4}$, as shown in Fig.\ \ref{Fig-2}, and in agreement with theoretically prediction by Khestanova \textit{et. al}. \cite{Khestanova2016} The obtained scaling in fact independently confirms the mechanism of the bubble formation through the interplay between the vdW and elastic forces. On the other hand, it shows that this rather simple effective substrate potential ($E_S$) can be used to effectively describe the adhesion at the interface between a graphene sheet and a substrate.

\begin{figure*}
\centering
\includegraphics[width=1.0\textwidth]{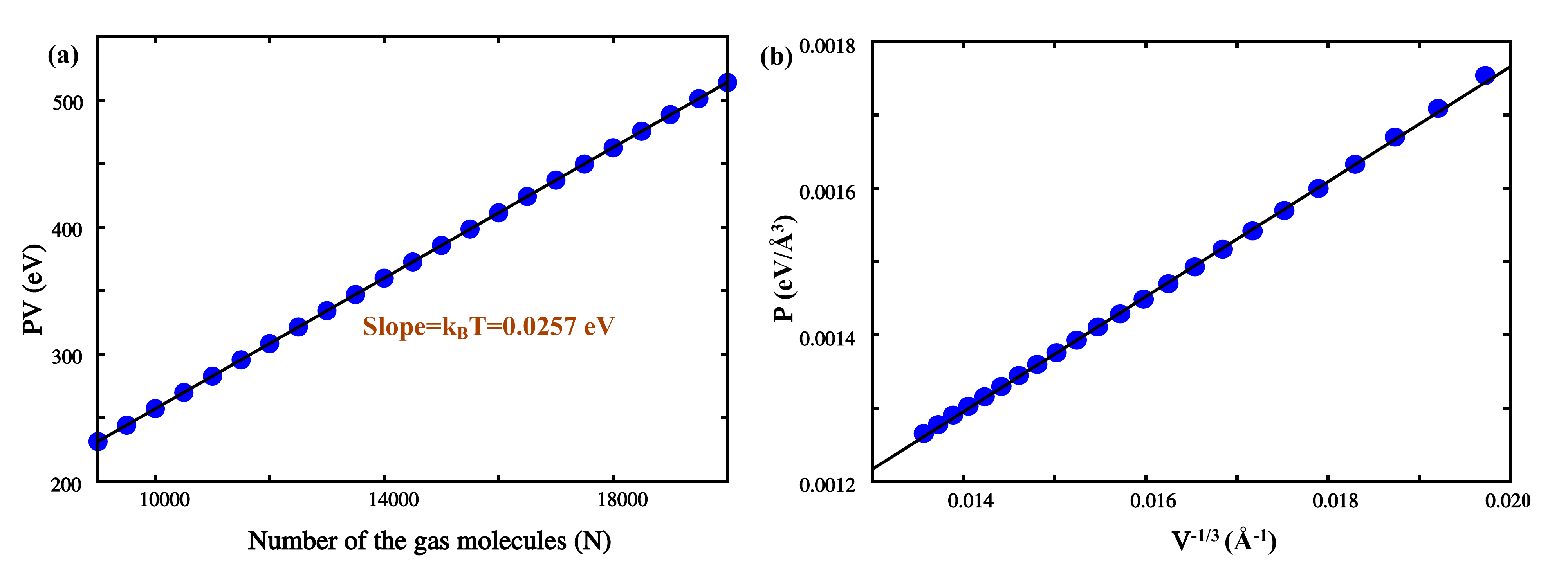}
\caption{Behavior of pressure $P$ and volume $V$ of the ideal gas trapped inside the bubble. (a) $PV$ versus number of gas molecules $N$. According to the ideal gas law $PV=NK_{B}T$; at room temperature the slope of the line is $K_B T=0.0257$ eV. (b) Plot of $P$ versus $V^{-1/3}$. Data points (blue) are fitted with a straight line, showing $P \sim V^{-1/3}$. }
\label{Fig-3}
\end{figure*}

Ideal gas is trapped inside the graphene bubble and in Fig.\ \ref{Fig-3}(a) we show its equation of state corresponding to temperature $T\simeq300$ K. From continuum elastic theory the pressure of the trapped ideal gas is predicted to scale as $P \sim V^{-1/3}$. This behavior is indeed observed in our numerical calculations, as shown in Fig.\ \ref{Fig-3}(b), which further corroborates the validity of this relatively simple theoretical approach for addressing the problem. The vdW pressure inside the bubble is of the order of 1 GPa and therefore has direct experimental consequences on the properties of the trapped materials.\cite{Xu2010,Algara-Siller2015,Yuk2012,Park2015,Wojcik2015}

\begin{figure*}
\centering
\includegraphics[width=1.0\textwidth]{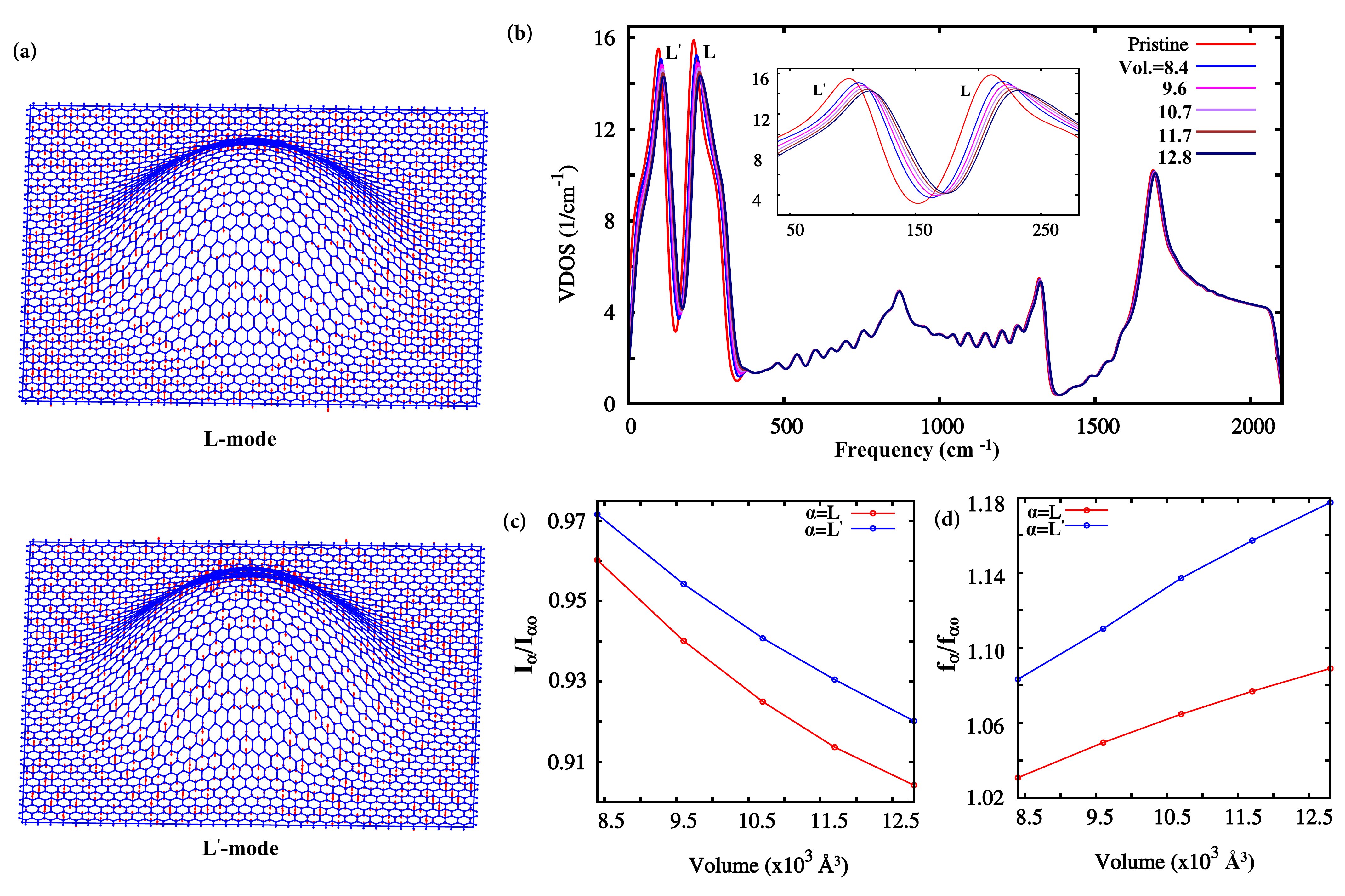}
\caption{Vibrational density of states (VDOS) and the profile of the prominent modes in the graphene nanobubble. (a) Profile of vibrational modes corresponding to $L$ and $L^{'}$ peaks, respectively. (b) VDOS of pristine graphene and samples with varying volume (in $10^{3}$\AA$^{3}$) of the ideal gas trapped in the bubble. Inset: Low-frequency peaks ($L$ and $L^{'}$) in VDOS zoomed in. (c) Relative decrease in the intensity of $L$ and $L^{'}$ modes at different volumes of the bubble. (d) Relative increase (blue shift) in the frequency of $L$ and $L^{'}$ modes at different volumes of the bubble.}
\label{Fig-4}
\end{figure*}

We now study an independent method for probing a graphene bubble via its characteristic "drumming" modes. More precisely, we here calculate the vibrational density of states (VDOS) of characteristic phonon modes of the graphene bubble. Out of all possible vibrational modes, most prominent are the out-of-plane $L$ and $L'$ modes, \cite{Jain2015} shown in Fig.\ \ref{Fig-4}(a). The presence of the bubble gives rise to a systematic shift in the position of the peaks in the VDOS corresponding to these two modes. As we can observe in Fig.\ \ref{Fig-4}(b), as volume of the bubble increases (pressure inside the bubble decreases) the frequency of both $L$ and $L'$ modes shifts towards higher values (blue shift). At the same time, this shift is correlated with a decrease of the intensity of these modes, as shown in Figs.\ \ref{Fig-4}(c) and \ref{Fig-4}(d). For a relative increase of the volume of the bubble by $\sim 50\%$, the relative decrease in the intensity of both modes is $\sim 5\%$, while the frequency of the $L$ mode increases by $\sim 5\%$ and that of $L'$ by $\sim 10\%$. These changes in the intensity and the frequency of these characteristic modes are therefore significant, and may be used as independent probes to measure shape and gas volume trapped in the nanobubbles in the inelastic electron tunneling and inelastic neutron spectroscopy.

\begin{figure*}
\centering
\includegraphics[width=1.0\textwidth]{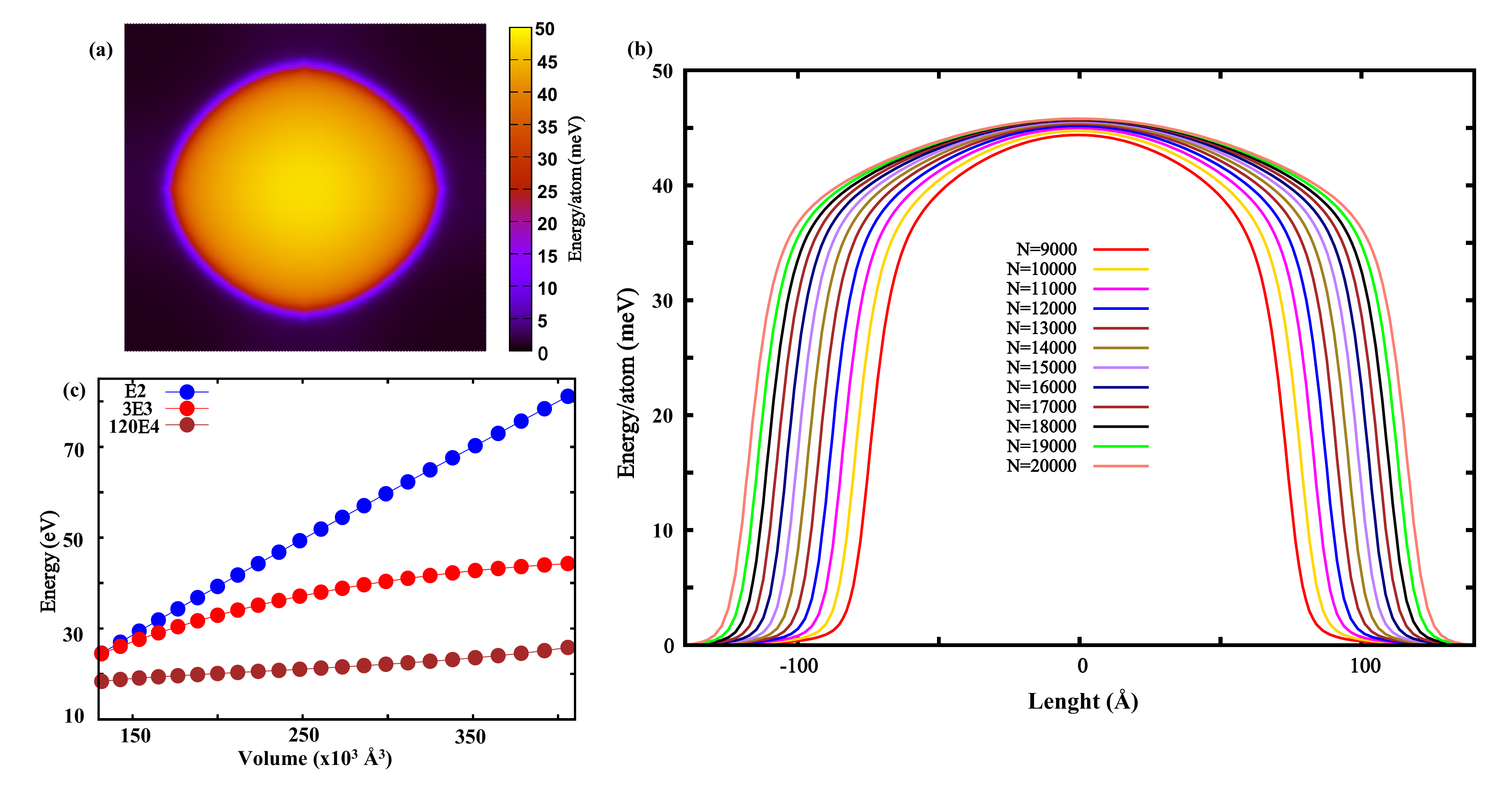}
\caption{Behavior of the elastic energy of a graphene layer with a trapped gas bubble. (a) Local energy plot of a sample with gas bubble ($N=18000$). (b) Profile of the energy distribution in the graphene layer as a function of the distance from the center of the bubble for different number of gas molecules trapped inside it. The center is determined by projecting the point of the maximum height onto the flat graphene plane. (c) Behavior of different elastic energy terms in eq.\ (\ref{eff-potential}) as a function of the bubble's volume: bond stretching ($E_2$), bond bending ($E_3$) and out-of-plane ($E_4$).}
\label{Fig-5}
\end{figure*}

We next study the elastic energy distribution of the graphene bubble. Elastic energy is directly related to strain, which is in turn important for strain engineering. We compute the elastic energy per atom using the combination of semi-empirical elastic potential eq.\ (\ref{eff-potential}) and adhesion energy ($E_S$) in eq.\ (\ref{eq-substrate}). In our samples we define a local energy per atom as follows: contributions due to two-body interactions are equally divided over the two interacting atoms, and contributions due to the three-body (angular) interactions are attributed to the central atom.\cite{Jain2016} Thus, the sum of the local energy over all atoms equals the total energy. This definition of local energy helps us to visualize the local degree of mechanical relaxation in the sample. The obtained profile of the local energy is shown in Fig.\ \ref{Fig-5}(a). Most of the elastic energy is concentrated on top of the bubble, and the energy is gradually decreasing as the edge of the bubble is approached. The shape of the energy profile in the radial direction does not completely follow the form of the bubble, as shown in Fig.\ \ref{Fig-5}(b), which may be important when engineering the strain-induced pseudo-magnetic fields.\cite{Levy2010} In particular, the energy and therefore strain at the top of the bubble remains rather constant when the number of trapped molecules increases from $N=9\times 10^3$ to $N=20\times10^3$. On the other hand, the distribution of the strain becomes more homogeneous as the number of trapped molecules increases, which results in less homogeneous pseudo-magnetic fields. In Fig.\ \ref{Fig-5}(c) we observe that as the volume of the bubble increases, the bond stretching terms contribute most to the total elastic energy of the bubble, and the strain arising from this type of elastic deformation is therefore dominant. Actually, when estimating the strain, the contribution coming from shearing and out-of-plane deformation can be, in a very good approximation, neglected, which should therefore simplify the calculations related to the pseudo-magnetic fields.

In conclusion, we have shown that a simple elastic semi-empirical potential combined with an effective substrate potential can accurately predict the universal shape of a gas bubble trapped between graphene and a substrate. With the help of numerical simulations we have shown that this universal behavior holds in small sized bubbles ($\sim 10$ nm) with the van der Waals pressure $\sim 1$ GPa. We find that the ratio of height and radius of a bubble scales as power law of the adhesion energy ($\sim K^{1/4}$). Furthermore, we have argued that the vibrational spectrum can be used as an independent probe of the shape of a trapped gas bubble. The obtained elastic energy distribution in the bubble shows that by manipulating the bond stretching and shearing, one can control strain engineering of the bubbles which is important for manipulating graphene's electronic, mechanical, adhesive and optical properties. The confinement pressure (vdW) and its effect should be taken into account in studies of various van der Waals heterostructures and can also be used to manipulate and modify the properties of trapped materials at the interfaces. In future, this computational approach can be useful to simulate other van der Waals heterostructures such as h-BN, Mo$_2$S, and WSe$_2$.

\newpage

\newpage

\section{Acknowledgments}
We acknowledge the support by FOM-SHELL-CSER program (12CSER049). This work is part of the research program of the Foundation for Fundamental Research of Matter (FOM), which is part of the Netherlands Organisation for Scientific Research (NWO).

\bibliography{Bubble}

\newpage

\begin{center}
\it { \textbf{Supplementary Online Material for ``Probing the Shape of a Graphene Nanobubble''}}

Sandeep K. Jain$^1$, Vladimir Juri\v ci\' c$^{2}$ and Gerard T. Barkema$^{3}$\\
\normalsize{$^1$}{Institute for Theoretical Physics, Universiteit Utrecht, Princetonplein 5, 3584 CC Utrecht, The Netherlands}\\
\normalsize{$^2$}{Nordita, Center for Quantum Materials, KTH Royal Institute of Technology,
and Stockholm University, Roslagstullsbacken 23, S-106 91 Stockholm, Sweden  }\\
\normalsize{$^3$}{Department of Information and Computing Science, Universiteit Utrecht, Princetonplein 5, 3584 CC Utrecht, The Netherlands }\\

\end{center}

\begin{figure*}
\begin{center}
\includegraphics[width=1.0\textwidth]{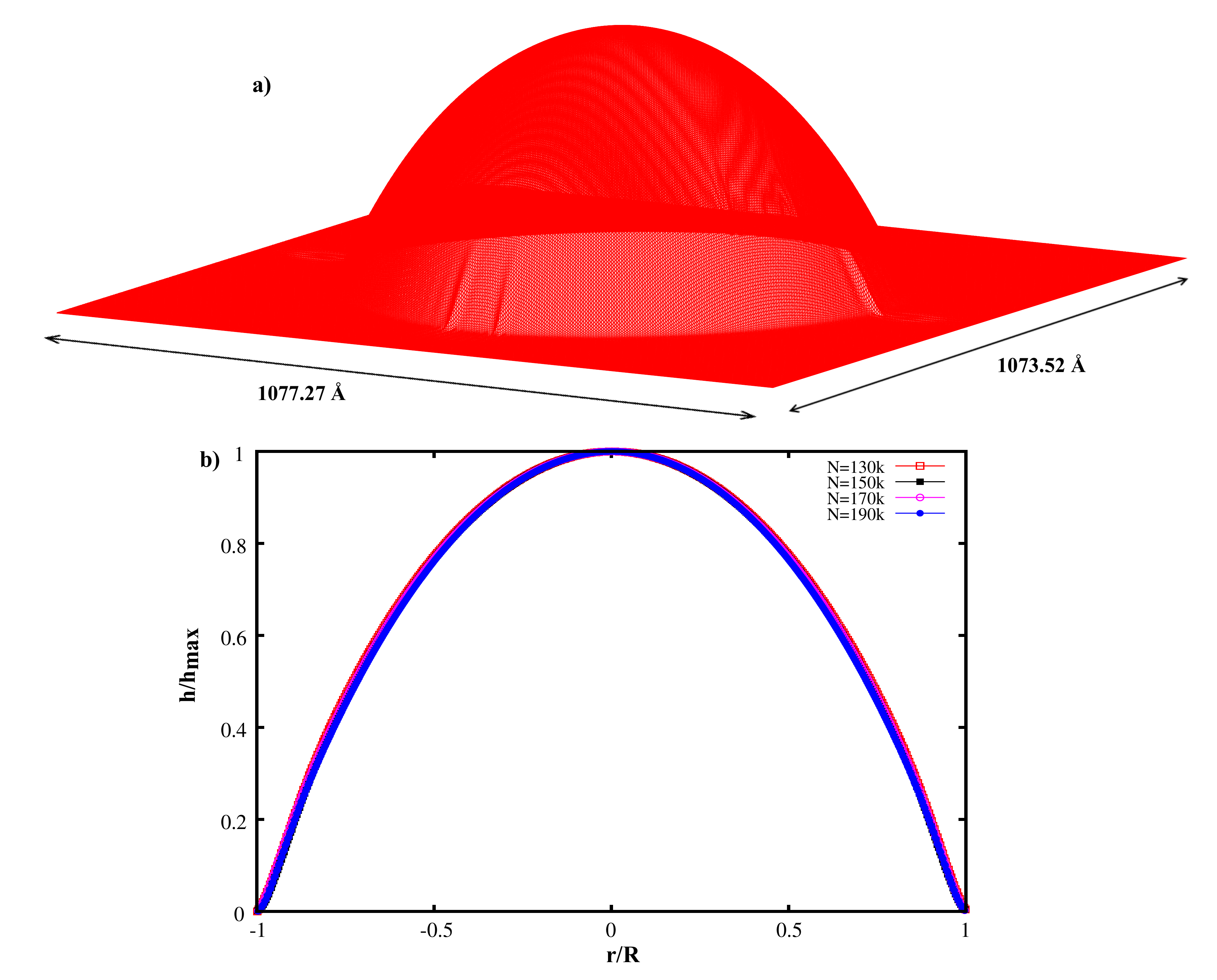}
\caption*{Figure S1: Profile of a large graphene bubble. (a) A gas bubble trapped between a graphene sheet with 441504 carbon atoms and a substrate. (b) The excellent collapse of the data of bubble's profile with varying sizes shows the universal shape of the bubble.}
\label{figure_S1}
\end{center}
\end{figure*}

\end{document}